\documentclass{emulateapj}
\def\stacksymbols #1#2#3#4{\def\theguybelow{#2}
        \def\verticalposition{\lower#3pt}
        \def\spacingwithinsymbol{\baselineskip0pt\lineskip#4pt}
        \mathrel{\mathpalette\intermediary#1}}
\def\intermediary #1#2{\verticalposition\vbox{\spacingwithinsymbol
        \everycr={}\tabskip0pt
        \halign{$\mathsurround0pt#1\hfil##\hfil$\crcr#2\crcr
                \theguybelow\crcr}}}
\def\lta{\stacksymbols{<}{\sim}{2.5}{.2}}
\def\gta{\stacksymbols{>}{\sim}{3}{.5}}

\shorttitle{TOTAL CLUSTER FEEDBACK ENERGY}
\shortauthors{MATHEWS \& GUO}

\begin{document}


\title{ESTIMATE OF THE TOTAL MECHANICAL 
FEEDBACK ENERGY FROM GALAXY CLUSTER-CENTERED BLACK HOLES: 
IMPLICATIONS FOR BLACK HOLE EVOLUTION, 
CLUSTER GAS FRACTION AND ENTROPY}

\author{William G. Mathews\altaffilmark{1} and 
Fulai Guo\altaffilmark{1}}

\altaffiltext{1}{University of California Observatories/Lick
Observatory,
Department of Astronomy and Astrophysics,
University of California, Santa Cruz, CA 95064 
mathews@ucolick.org}

\begin{abstract}
The total feedback energy injected into hot gas in galaxy clusters by
central black holes can be estimated by comparing the potential energy
of observed cluster gas profiles with the potential energy of
non-radiating, feedback-free hot gas atmospheres resulting from
gravitational collapse in clusters of the same total mass.  Feedback
energy from cluster-centered black holes expands the cluster gas,
lowering the gas-to-dark matter mass ratio below the cosmic
value.  Feedback energy is unnecessarily delivered by radio-emitting
jets to distant gas far beyond the cooling radius where the cooling
time equals the cluster lifetime.  For clusters of mass $4-11\times
10^{14}$ $M_{\odot}$ estimates of the total feedback energy,
$1-3\times 10^{63}$ ergs, far exceed feedback energies estimated from
observations of X-ray cavities and shocks in the cluster gas, energies
gained from supernovae, and energies lost from cluster gas by
radiation.  The time-averaged mean feedback luminosity is comparable
to those of powerful quasars, implying that some significant fraction
of this energy may arise from the spin of the black hole.  The
universal entropy profile in feedback-free gaseous atmospheres in NFW
cluster halos can be recovered by multiplying the observed gas entropy
profile of any relaxed cluster by a factor involving the gas fraction
profile.  While the feedback energy and associated mass outflow in the
clusters we consider far exceed that necessary to stop cooling inflow,
the time-averaged mass outflow at the cooling radius almost exactly
balances the mass that cools within this radius, an essential
condition to shut down cluster cooling flows.
\end{abstract}

\vskip.1in

\section{Introduction}

Massive black holes in the cores of group and 
cluster-centered galaxies become active 
when a tiny mass of cluster gas is accreted, 
causing large amounts of ``feedback'' energy 
to be deposited in the surrounding cluster gas.
The energy provided by occasional feedback events,  
in the form of jets containing supra-thermal and 
relativistic particles,  
increases the entropy of both nearby and very 
distant cluster gas.
Feedback jets create 
expanding, shock-driving cavities in the hot cluster gas, 
increasing its entropy and delaying its flow toward  
the black hole by radiation losses. 
In the absence of feedback, the loss of entropy 
by thermal X-radiation is accompanied by a slow 
cooling inflow toward the black hole which, 
if unabated over time,  
would concentrate a mass of centrally cooled gas in or 
near the central black hole far exceeding 
limits set by stellar velocities and other observations.
For example, intermittent cavities formed from 
feedback events of energy $10^{59}$ ergs 
every $2\times 10^8$ years at 10 kpc can arrest the 
currently observed gas density and temperature profiles 
(and low central cooling) 
in the Virgo cluster for $\sim3$ Gyrs (Mathews 2009). 
Including more distant feedback events 
at $50$ kpc can maintain the observed gas profiles 
and low cooling rate for several 
additional Gyrs. 
Entropy and cosmic rays 
delivered to the cluster gas by feedback episodes cause  
gas to flow out in the cluster potential, offsetting 
the cooling inflow.

In general, however, central black holes are unable 
to provide the optimal or minimum feedback energy that must 
be deposited at every radius in the cluster gas 
just sufficient to shut down black hole accretion 
of locally cooling gas.
Instead, black holes typically over-react 
in a clumsy fashion, depositing 
much more energy than the minimum required, 
much of it in distant regions of the cluster where the radiative 
cooling time exceeds the age of the cluster. 
After a few $10^8$ yrs following a feedback event 
most of the feedback energy converts to potential energy,
as the entire gaseous cluster atmosphere 
adiabatically expands outward
(Mathews \& Brighenti 2008). 
Consequently, the increased cluster gas entropy is 
necessarily related to a reduction of the cluster gas density 
relative to the local dark matter.

We discuss here an approximate estimation of the total
feedback energy received by cluster gas during the 
cluster lifetime by comparing gas potential 
energy profiles in observed clusters with that of idealized 
gas density distributions resulting from 
``adiabatic'' gravitational collapse into the cluster halo 
in the absence of radiative cooling and associated feedback.
We show that this energy, $\sim10^{63}$ ergs, 
far exceeds the energy lost by radiation during the cluster 
lifetime and consequently the minimum energy required 
merely to stop the cooling inflow. 
The collective energy from all supernovae also provides 
a negligible fraction of the total feedback energy.  
Since only 10 percent of the hot baryonic gas in massive clusters 
cools to form stars, star formation can also be ignored 
in our estimate of the global energetics 
of the cluster gas where we seek an overall 
accuracy of $\sim25$\%.

Estimates of the total feedback energy from cluster-centered
black holes are possible for two reasons. 
First, the potential energy of 
cluster gas in hydrostatic equilibrium 
in a fixed dark halo potential 
can be found by integrating outward from the cluster center.
Second, to a good approximation,  
the formation of dark halos and their gravitational potential 
proceeds from the inside outward as the 
size of the virialized 
region in the dark halo increases with time. 
In view of this latter point, it is possible to estimate 
the increase in potential energy of the cluster gas 
due to feedback without knowing when the feedback occurred.

We begin by estimating the current gas density, temperature, 
mass, entropy and potential energy profiles 
expected at zero redshift in the absence 
of radiative losses, star formation, and black hole feedback, 
referred to as the ``adiabatic'' cluster atmosphere.
Then, assuming hydrostatic equilibrium, 
potential energy profiles are estimated from 
gas distributions in observed clusters having the same total mass.
When comparing the difference between the
potential energy evaluated at the same mass of cluster gas 
with and without feedback, 
we find that hot gas in the idealized 
adiabatic atmosphere must expand significantly  
to resemble cluster gas profiles currently observed. 
The total feedback energy associated with this expansion, 
$\sim 1-3\times 10^{63}$ ergs, comfortably exceeds 
energies of the most powerful known individual feedback events.
Central to our feedback energy estimate is the 
inverse relationship between increasing cluster gas entropy
and decreasing gas fraction, the ratio of gas to total 
cluster densities, which is lowered by a global expansion 
of the cluster gas.

If this feedback energy is released during 
periods of radiatively efficient central accretion 
with $\sim 0.1$ of the accreted
mass returned as feedback energy to the cluster gas,
we find that the final black hole masses 
in large clusters would exceed those observed.
Alternatively, the feedback energy may be provided 
from the rotational energy of 
rapidly spinning central black holes.

\section{Estimate of Total Feedback Energy}

A key element in our estimate is an assertion that 
adiabatic (non-radiating, non-feedback) gaseous cluster
atmospheres can be fit with the same 
properly normalized NFW profile as the dark matter. 
Among the many computations of galaxy cluster formation 
that include both baryons and dark matter, 
surprisingly few have analyzed in detail the 
deep similarity between the final adiabatic gas structure 
and the radial NFW distribution of dark matter. 
Notable exceptions to this are 
the density and entropy profiles 
in dark halos and adiabatic gas 
described in the cosmological cluster formation 
calculations of Eke, Navarro \& Frenk (1998) and 
in particular Faltenbacher et al.  (2007).
The cluster gas entropy  
can be characterized with $S_g = \sigma^2/\rho^{2/3}$
where $\rho$ is the gas density  
and $\sigma = [3 k T/(\mu m_p)]^{1/2}$ is the thermal velocity dispersion.
By analogy, Faltenbacher et al. define a corresponding 
dark matter entropy $S_{dm} = \sigma^2/\rho^{2/3}$ 
for which $\rho$ is the dark matter density and 
$\sigma$ is the 3D velocity dispersion of collisionless
dark matter particles.
The detailed cosmological cluster calculations of 
Faltenbacher et al. (2007) using GADGET2 reveal that,  
apart from a small central gaseous core, 
the gas density and 
entropy profiles in the adiabatic case share identical, 
appropriately scaled NFW profiles regardless of cluster mass.
Beyond the central core where $S_g > S_{dm}$,
both $S_g$ and $S_{dm}$ vary as powerlaws, 
$ \propto r^{1.21}$, but with slightly different 
normalizations, $S_g/S_{dm} = 0.71 \pm 0.18$. 
However, as discussed by Faltenbacher et al., 
the final zero redshift gas configuration 
in the gas contains residual 
subsonic macroscopic velocities which, if damped, would bring 
the gas and dark matter entropies into near perfect 
agreement, $S_g/S_{dm} \approx 1$. 
We adopt the assumption, implicit in the discussion of 
Faltenbacher et al., that the two entropies 
are in fact equal but the gas remains undamped 
because the appropriate  
physical damping mechanisms are absent from GADGET2. 
The artificial viscosity in this code that damps the accretion 
shock may require an unphysically long time to 
damp the residual subsonic velocity field.
Assuming $S_g \approx S_{dm}$ is also more consistent with 
the smaller $\sim5$\% component of turbulent energy found 
for relaxed clusters in recent cosmological simulations using 
the ENZO code (Vazza et al. 2011). 
Finally, cluster virial masses are determined 
from observed gas pressure profiles without allowing for 
undamped kinetic energy and its associated pressure. 
This assumption (which underestimates the virial mass)  
is also consistent with $S_g \approx S_{dm}$.
While it is comforting that the radial profiles of 
adiabatic gas and dark matter 
are nearly identical, this may nonetheless be surprising 
due to the very different nature of their dissipative 
mechanisms\footnotemark[2]. 
Of more relevance to our discussion is the similar NFW shape 
of adiabatic gas and dark matter density profiles.
The assumption $S_g \approx S_{dm}$ ensures that 
the adiabatic gas has experienced the same dissipative history
as the dark matter. 

\footnotetext[2]{
In adiabatic cluster simulations
gases of different entropies evidently mix in the cluster core,
raising the total gas entropy, 
particularly in grid-based computations.
However, 
beyond the core we expect the density of gas and dark matter 
to share the same appropriately normalized NFW profile. 
This profile similarity has been verified in many 
calculations including the Santa Barbara cluster, 
an average of 12 different structure formation 
codes using identical cosmologies 
and initial conditions (Frenk et al. 1999).
In more recent adiabatic simulations 
the effective dark matter temperature profile
$T_{dm}(r) = (\mu m_p/3 k)\sigma^2$ 
has been shown to be identical (within 10\%) to the gas temperature
profile $T(r)$ 
(e.g. Host et al. 2009; ZuHone 2011). 
}

The radial dark matter NFW distribution is shaped by all 
entropy-producing dissipations that occurred 
during both smooth accretion of diffuse dark matter as well as 
inhomogeneous accretion of smaller groups, clusters and filaments 
that merged into the final cluster potential.  
The entropy increases that accompanied the formation 
of these smaller structures are also embedded in the 
final NFW structure.
Furthermore, 
the results of Faltenbacher et al. (2007) indicate that
all dissipative information about the cluster merger 
history is also encoded in the NFW gas density profiles 
in idealized adiabatic baryon atmospheres 
formed by gravity alone. 

When computing purely gravitational (adiabatic) collapse in 
cosmological calculations without radiation or feedback, 
a controversy has arisen in recent years concerning
differences in the dissipation 
between dark matter and baryons in cluster cores 
where $S_g > S_{dm}$   
(recently reviewed by Borgani \& Kravtsov 2009; 
Springel 2010; Vazza 2011).
In mesh-based calculations the baryons are found to have 
large, well-resolved 
central density cores which are larger than those computed 
with smoothed particle hydrodynamics.
Possible origins for this discrepancy have been proposed 
and discussed by 
Mitchell et al. (2009) and Vazza (2011). 
For our purposes here we assume two limiting 
adiabatic gas density 
profiles following pure gravitational baryonic 
collapse into dark halos: core and no core. 

\subsection{Adiabatic cluster gas atmospheres without cores}

Consider first the ``no core'' case in which baryons 
and dark matter suffer the same dissipation so the 
post-collapse gas density profile is identical to that of 
the total density but scaled down  
by the universal baryon fraction $f_b = 0.17$, 
i.e. $\rho = f_b\rho_t$ 
(e.g. Faltenbacher et al. 2007) 
where $\rho_t$ is the total density dominated 
by dark matter. 
We seek the idealized radial structure of cluster gas 
formed without feedback or radiative losses 
and which has evolved to the current time.
The temperature profile can be found by integrating the 
equation of hydrostatic equilibrium
\begin{equation}
{d \theta \over dr} = 
\theta \left(-{1 \over \rho}{d \rho \over  dr}\right) - g
\end{equation}
where $\theta = kT/\mu m_p$ and $g = GM(r)/r^2$. 
The self-gravitation of the gas 
is implicitly included in the total mass $M(r)$ that 
includes both dark matter and gas.
The total matter density and mass profile 
in a cluster with virial mass $M_v$ are 
given by the usual NFW relations,
\begin{equation}
\rho_t = {\rho_c \delta_c \over y(1+y)^2}~~~{\rm and}~~~M(r) 
= M_v{f(y) \over f(c)}
\end{equation}
where $c = y/(r/r_v)$ is the concentration and 
\begin{equation}
f(y) = \ln(1+y) - y/(1+y)~~~\delta_c = \Delta_c c^3/3 f(c).
\end{equation}
The radius where the local cluster density is 
$\Delta$ times larger than the critical density 
$\rho_c = 3 H^2/8\pi G = 9.24\times 10^{-30}$ g cm$^{-3}$ is 
\begin{equation}
r_{\Delta} = (3 M_v / 4 \pi \Delta \rho_c)^{1/3}
\end{equation}
and the virial radius is 
$r_v = r_{\Delta_c}$ 
where $\Delta_c = 178(\Omega_m)^{0.45} = 103$ 
(Eke, Navarro, Frenk 1998) when $\Omega_M = 0.3$. 
For nearby clusters we assume 
$H = 70$ km s$^{-1}$ Mpc$^{-1}$.
Finally, we adopt a convenient concentration-$M_v$ relation  
$c = 9 (M_v/10^{14}M_{\odot})^{-0.172}$ 
for low redshift clusters 
(Buote et al. 2007).


In our integrations of equation (1) 
using the observed gas density 
we consider two 
clusters each of which are composites of two very similar 
clusters selected from the sample of nearby relaxed clusters
observed with the {\it Chandra} telescope 
by Vikhlinin et al. (2006).
For our purposes we choose massive clusters 
in which the full impact of feedback energy can be absorbed 
and which do not presently have huge cavities that 
dominate and disrupt the cluster gas profiles observed.
Of particular interest are the NFW parameters 
determined for these clusters 
and values of the observed cluster gas fraction
$f_g = \rho/\rho_t$ 
plotted in Figures 3-14 of Vikhlinin et al.
To avoid being distracted by
spurious observational errors
and to improve the quality of our estimates,
we combine the mean properties of two pairs of clusters 
that share nearly the same virial mass: 
$\langle A133 + A383\rangle$ and
$\langle A478 + A1413\rangle$.
The two averaged clusters are 
subsequently referred to as composite clusters 1 and 2 
respectively.
Relevant observed NFW properties of all four clusters 
and the the two composite clusters are listed in Table 1.
Each composite cluster is chosen to have 
values of $r_{500}$ and $M_{500}$ that are averages of 
the two observed clusters.
Composite clusters 1 and 2 have virial masses 
$M_v = 4.34 \times 10^{14}$ and 
$M_v = 11.3 \times 10^{14}$ $M_{\odot}$ respectively.

Our first integrations of equation (1) are for adiabatic ``no core'' 
versions of the two composite clusters 
using $\rho(r) = f_b \rho_t(r)$ for the gas density.
We choose an initial temperature 
at some very small cluster radius and seek solutions for 
which the gas entropy $S = \theta/\rho^{2/3}$ varies like a 
self-similar power law at large radius $S \propto r^{1.2}$, 
similar to entropy profiles 
in detailed adiabatic gravitational collapse 
computations (e.g. Tozzi \& Norman 2001; Voit et al. 2005).
This is a well-defined initial value problem 
with unique solutions.
By varying the initial temperature,
an exact value and solution 
can be found for which $S \propto r^q$ 
where $q \approx 1.2$ remains constant over a large range 
of cluster radius near the virial radius and beyond. 
The value of $q$ emerges naturally from the solution 
as an eigenvalue that is not imposed in advance.

Solid lines in Figures 1 and 2 show adiabatic post-collapse
profiles of gas temperature, entropy and density respectively
for composite clusters 1 and 2
found by integrating equation (1).
The hydrostatic post-collapse cluster gas 
atmospheres in Figures 1 and 2 have a broad temperature maximum near 
$r/r_v = 0.1$ and distant entropy profiles 
$S \propto r^{1.3}$ that are very similar to the 
Tozzi-Norman computation.
Our simple approximation closely resembles 
sophisticated cluster gas 
profiles computed in detailed cosmological simulations 
that include mergers of smaller bound systems and gas 
inflowing in filaments (e.g. Borgani \& Kravtsov 2009).
Gas and dark matter in these 
merging sub-systems and filaments have already experienced 
some dissipation when they enter the cluster virial radius. 
Apart from this we consider no other type of 
hypothetical ad hoc pre-heating.

\subsection{Observed cluster gas atmospheres without cores}

Next we perform similar integrations using the observed 
cluster gas density profiles $\rho_{obs}(r)$,   
approximately including the gas self-gravity 
in the total NFW mass distribution.
The data points shown with open circles 
in Figures 1 and 2 show observed gas densities 
found from $\rho_{obs}(r) = f_g(r)\rho_t(r)$ 
for both clusters in each composite cluster 
at radii for which $f_g$ is determined by Vikhlinin et al. 
In addition we show (with filled circles) several 
additional approximate values measured 
directly from the $\rho_{obs}(r)$ profiles 
plotted by Vikhlinin et al.
(Closed and open circles have sizes 
roughly comparable to observational errors.)

The mean observed gas density profiles of our two composite 
clusters are fit with an analytic curve 
\begin{equation} 
\rho_{obs}(r) = {f_b \delta_c \rho_c \over
y^{1-\alpha} (y_0 + y)^{2+\alpha}}
\end{equation}
where $y = c(r/r_v)$. 
By design, $\rho_{obs}(r)$ asymptotically approaches 
$f_b\rho_t(r) = 0.17\rho_t(r)$ as the radius continues beyond the 
observed region, i.e. $r \gta 0.5r_v$. 
As the cluster gas conserves baryons 
during feedback expansion, regions of 
$f_g < f_b$ in the cluster gas observed within $\sim0.5r_v$
must be compensated by regions of $f_g > f_b$ in more 
distant cluster gas. 
However, due to the increased volume available 
in the outer regions of the clusters, 
we expect the excess $f_g - f_b$ to be much 
less than $f_b$ and difficult to observe.
Profiles for $\rho_{obs}(r)$ are shown with dotted lines 
in the bottom panels of Figures 1-4 where 
we take $y_0 = 1.8$ and 1.5 for cluster 1 and 2
respectively and $\alpha = 0.05$ for both composite clusters.

For the observed hot gas atmospheres in clusters 1 and 2  
equation (1) is solved in the same manner as before 
but with $d\ln \rho_{obs}/dr$ on the right hand side.
The corresponding hydrostatic 
temperature $T(r)$ and entropy $S(r)$ 
profiles are shown with 
dotted lines in the upper and central panels of Figures 1 and 2
for each composite cluster.
Both the temperature maximum and the somewhat 
flatter entropy profile that asymptotically approaches 
the adiabatic profile resemble typical cluster observations 
(e.g. Vikhlinin et al. 2006; Pratt et al. 2010). 
The total bolometric X-ray luminosity within the virial 
radius 
$L_x(r_v) = (3.83 \times 10^{44}$, $1.58 \times 10^{45})$ 
erg s$^{-1}$ respectively for composite 
clusters (1,2) are consistent with the $L_x - M$
scaling relation for observed clusters (e.g. Maughan 2007).

While the two cluster atmospheres plotted in
Figures 1 and 2 -- adiabatic and observed -- 
are final zero-redshift gas distributions, 
we can also regard them  
as initial and final configurations
of the cluster gas before and after feedback
energy is deposited.
Of particular interest is the potential energy of the 
gas at various radii in the two 
atmospheres that enclose the same integrated gas mass $M_g(r)$. 
The potential energy of gas in the cluster is 
\begin{equation}
PE(r) = \int_0^r \phi(r) \rho 4 \pi r^2 dr
\end{equation}
where the NFW potential 
\begin{equation}
\phi = -{G M_v \ln(1 + y)  \over r f(c)} 
\end{equation}
also satisfies $g = - d\phi/dr$. 
In evaluating the potential $\phi(r)$ we again approximate 
the self-gravity of the gas by using the total 
cluster mass, 
and we ignore the small adiabatic expansion of the 
dark matter as gas moves out in the cluster potential. 
The NFW potential is expected to be valid 
within about twice the virial radius 
(Cuesta et al. 2008; Tavio et al. 2008).

Note that the potential energy profile 
is determined by an 
integration out from the cluster center. 
Furthermore, to a good approximation, 
the cluster dark matter density and potential 
profiles remain fixed with time as 
the the size and mass of the virialized region 
increases with time,
i.e. the mass distribution in virialized dark halos 
forms from the inside out.
The inside-out character of halo formation is apparent 
for example from Figure 1 of Diemand, Kuhlen, Madau (2007)
who plot the accumulated density within many 
Lagrangian mass zones with time as a dark halo grows 
in size and mass. 
In view of this, our estimate of the change in $PE$ 
due to non-gravitational feedback,  
assuming an unchanging total NFW mass distribution, 
is independent of the time when the feedback occurred.
As the dark halo grows, 
the global outflow of cluster gas driven by feedback 
energy can relocate the accretion shock 
beyond the instantaneous virial radius. 
But we do not consider extremely distant 
feedback events ($r \ga 1.5-2r_v$) that energize gas 
before it reaches the accretion shock 
and where the NFW potential may no longer apply.
This restriction on feedback seems reasonable, particularly 
since the virial radius continuously increases with time 
and it is consistent with the gradual radial 
increase in $f_g$ toward $f_b$ observed in cluster gas.

Radii and potential energy $PE$ values at three locations  
in the ``no core'' adiabatic and observed solutions  
having the same $M_g(r)$ are listed at the top of Table 2 
for clusters 1 and 2.
It is seen that the radius 
that encloses all the gas within the virial radius 
in the pre-feedback adiabatic cluster
increases by about a factor 1.9 (cluster 1) or 
1.5 (cluster 2) 
as the cluster gas expands due to feedback energy. 
This expansion has been verified by George et al. 
(2009) who observe a galaxy cluster with 
virialized, X-ray emitting gas well beyond the virial radius.
The increase in potential energy due to feedback 
when integrated to the same gas mass, 
shown as $\Delta PE$ in Table 2,  
is huge, $1-3\times 10^{63}$ ergs, 
greatly exceeding the most energetic known 
individual bipolar feedback events $\lta 10^{62}$ ergs 
(McNamara et al. 2005; Guo \& Mathews 2010a), 
which are sufficient to convert cool-core  
to non-cool-core clusters (Guo \& Mathews 2010b).
The mean luminosity of this enormous feedback energy, 
if spread over a typical cluster lifetime $t_{cl}\sim 7$ Gyrs, is 
$L_{fb} \approx 4\times 10^{45} (|PE|/10^{63}{\rm ergs})$ 
erg s$^{-1}$, comparable to continuously active quasars. 

\subsection{Adiabatic and observed cluster gas atmospheres 
with cores}

In a second series of similar calculations, we consider 
adiabatic collapse atmospheres in which the baryons 
experience an additional central dissipation that produces
a density core. 
For this purpose, we adopt the recent mesh-based purely 
gravitational cluster collapse computations of 
Vazza (2011). 
We create a core from the no core density profile by 
flattening the central gas density 
so that $d\log \rho /d\log r \approx -2$
at $r/r_{v} \approx 1.4$ as in  
Figure 3 of Vazza (2011). 
However, to satisfy the stability requirement of radially 
increasing entropy, the gas density in the core 
cannot be perfectly flat.
We find positive $dS/dr$ when the core gas density 
gradually decreases as $\rho \propto r^{-0.3}$ and this 
solution is shown as a solid line 
in the bottom panels of Figures 3 and 4. 
In our approximation baryons in the core region are simply 
removed without a corresponding increase in $f_g$ beyond 
the core region. 
Corrections of this sort that must occur for baryon conservation, 
and which are not apparent in the mesh-based calculations, 
are not important for our estimate here since the 
baryon fraction in the cored adiabatic atmosphere 
is essentially $f_b = 0.17$ at large cluster radii. 
Finally, we assume that adiabatic cored density profiles 
are a universal function, scaling with $r/r_v$ for clusters 
of different virial masses. 
Performing similar integrations of the hydrostatic equation (1), 
temperature and entropy profiles corresponding 
to the cored density profile 
are shown as solid lines for composite clusters 1 and 2 
in the upper two panels of Figures 3 and 4 
where it is seen that the entropy has a slowly 
sloping entropy ``floor'', as expected.

Repeating the same procedure as before, 
three pairs of equal gas mass $M_g(r)$ locations in the 
two cluster gas atmospheres -- adiabatic ``core''
collapse and observed -- are listed at the bottom of Table 2.
When baryonic cores are present 
the potential energy increases $\Delta PE$ are only slightly 
less than if the core is completely ignored.

\subsection{Further observational implications}

Both of the starless adiabatic cluster atmospheres
we consider -- with and without cores -- have local gas fractions
that agree with the cosmic baryon value $f_b = 0.17$
near $r_{v}$.
However, the large expansion of cluster gas caused by
feedback reduces the local gas fraction throughout
the observed post-feedback cluster.
The observed gas fraction
at the most distant density observations
near $r \approx 0.4r_{v}$ is $f_g = 0.082$,
only about half the cosmic value.
This shortfall is typical of all massive clusters.
The average value of the gas fraction at $r_{500} \approx 0.5r_{v}$
for clusters having mean temperatures greater than 4 keV
is $f_g(r_{500}) \approx 0.12 \pm 0.02$
(McCarthy, Bower \& Balogh 2007).
When feedback-energized cluster gas expands out beyond $r_{500}$,
baryon conservation requires that $f_g > f_b$
somewhere beyond $r_{500}$,
although baryon excesses have not yet been observed.
Nevertheless, it is possible that
nearly baryonically closed systems in which
$\Delta PE << |PE(r_{v})|$ and
$f_g(r_{v}) \approx 0.17 $
may exist in some fossil groups where
the total mass of gas cooling near the central black hole
is much less than in the more massive clusters considered
here (Mathews et al. 2005).

The final pair of entries for each of the four cluster
atmospheres in Table 2
are designed for the majority of current
cluster observations that extend only to
$\sim r_{500} \approx 0.5r_{v}$.
Values of $\Delta PE(r_{ad})$
with $r_{ad} \approx 0.3r_v$ in Table 2
represent the minimum feedback energy consistent
with clusters observed to $\sim 0.5r_v$.
This minimum feedback
energy -- about $3-13\times 10^{62}$ ergs from Table 2 -- would
be correct if $f_g$ increases abruptly to
$f_b = 0.17$ just beyond the outermost gas density observed.
While a few recent cluster X-ray observations
with the {\it Suzaku} telescope
extend to the virial radius and beyond
(e.g. George et al. 2009),
these data may be more difficult to interpret.
Detailed calculations of the cosmological
evolution of massive clusters
often indicate significant undamped subsonic gas
velocities beyond about
$r_{500} \approx 0.5r_{v}$, which may degrade the assumption
of hydrostatic equilibrium (Evrard et al. 1996).
Baryonic clumps or compressions occur in this same outer region,
causing the observed gas density to appear too high and the
entropy too low (Nagai et al. 2007a,b).

\subsection{Influence of stellar baryons}

In our estimates of the feedback energy 
we have ignored the small fraction of 
baryons in our clusters that convert to stars.
Nevertheless, 
it is interesting to compare the mass of stellar baryons 
to the total baryonic gas mass that flows 
out at large cluster radii due to feedback energy.
The mean cluster stellar mass fraction $f_*$ 
decreases with cluster mass 
as $f_* \propto M_v^{-0.55}$ (Andreon 2010) but this 
variation has considerable cosmic scatter. 
For our clusters 1 and 2 with mass 
$\log M_{200} = 14.56$ and 14,96 in $M_{\odot}$, 
the stellar fraction, $f_* \approx 0.008-0.015$ 
(from Figure 7 of Andreon 2010), is dominated by 
cosmic scatter. 
Adopting a mean value for this range of cluster mass, 
$f_* \approx 0.010$, the total stellar mass within 
$0.5r_v$ in clusters (1,2) is 
$(0.26,0.65)\times 10^{13}$ $M_{\odot}$.
By comparison, from Table 2 the total mass flowing 
out beyond $0.5r_v$ for (core,no core) versions of 
cluster 1 is $(1.823,1.983)\times10^{13}$ $M_{\odot}$ 
and $(3.69,3.91)\times10^{13}$ $M_{\odot}$ for cluster 2.
The fraction of outflowing gas mass consumed by star formation, 
$\sim 0.13-0.17$, 
is another small correction that we neglect here.

\subsection{Specific feedback energy}

Our values of 
$\Delta PE(r_v) = (8.34,29.1)\times 10^{62}$ ergs 
for 
cored clusters (1,2) translate into mean specific 
feedback energies 
$\Delta PE/(0.17 M_v) = (5.65,7.57)\times  10^{15}$ erg gm$^{-1}$ 
when applied to the total gas mass within $r_v$
in the original adiabatic clusters 
that extends out to $(1.5,1.8)r_v$  
in the post-feedback atmosphere. 
The specific feedback energy can also be expressed as 
$(3.6,4.8)$ keV particle$^{-1}$.
These values appear comparable to those estimated 
from the $L_X - T$ scaling relation 
for clusters by Wu, Fabian \& Nulsen (2000),
$\sim1-3$ keV particle$^{-1}$. 
However, Wu et al. only consider particles in 
the mass of observed gas within $r_{200}$; 
our gas masses are larger by (2.0,1.8) so our estimated 
feedback energies are more than twice as large 
as those of Wu et al.
By comparison, our specific feedback energies are 
considerably less than those of McCarthy, Bower, \& Balogh 
(2007), $\gta 10$ keV particle$^{-1}$, 
probably because they use an extrapolation 
procedure for the observed gas density $\rho_{obs}(r)$ 
quite different from our equation (5).

\subsection{Energetics of radiation loss and supernovae}

We assume that the
energy and entropy lost by radiation can be ignored.
To verify this, consider
the total energy radiated by the two composite clusters
within the cooling radius
$E_{rad} \approx L_x(r_{cool})t_{cl}$
during the cluster lifetime $t_{cl} = 7$ Gyrs.
For composite clusters (1,2) we find
$r_{cool} = (97.7,119.8)$ kpc,
$L_x(r_{cool}) = (1.36,4.96)\times 10^{44}$ erg s$^{-1}$,
and $E_{rad} = (0.30,1.1)\times 10^{62}$ ergs.
The total radiated energy for both clusters is a small
fraction of the change in potential energy $\Delta PE$ listed in
Table 2, particularly those values of
the total $\Delta PE(r_v)$
evaluated at the virial radius in the adiabatic solutions.
Radiation losses, while not a major factor in the overall
energy budget for our clusters, are nevertheless essential in
driving the central cooling accretion that creates
the feedback energy.
The cooling radius ($\log (r_{cool}/r_v) \approx -1.3$)
is small compared to the more extended regions in
the bottom panels of Figures 1-4
where the cluster density is observed to be significantly
below the maximum cosmic baryon density, $\rho_{obs} < f_b\rho_t$.
Feedback energy from cluster-centered black holes
has obviously caused huge outflows, removing cluster gas
from regions far beyond the cluster cooling radius.

Feedback from Type II and Ia supernovae can also be ignored.
For a single-population Salpeter IMF,
about 0.007 Type II supernovae events occur
per solar mass of gas formed into stars.
Consider the total mass of gas observed
in the composite observed clusters (1,2),
extrapolated to the virial radius,
about $M_g(r_v) = (4.7,13.9) \times 10^{13}$ $M_{\odot}$.
If 10\% of the baryons in this mass forms into stars,
the total feedback energy from SnII, each of energy
$10^{51}$ ergs, is of order
$E_{SnII} \approx (0.3,0.9)\times 10^{62}$ ergs which is
substantially less than
the $\Delta PE(r_v)$ in Table 2 evaluated
at radius $r_v$ in the pre-feedback gas.
If the average iron abundance in cluster gas is about 0.3 solar,
the total mass of iron within $r_v$ in the clusters is roughly
$0.3\times 0.0017\times M_g(r_v) = (2.4,7.1)\times 10^{10}$
$M_{\odot}$ where 0.0017 is the fraction of iron by mass
in the solar photosphere.
If all the iron is created in Type Ia supernovae,
each having energy $10^{51}$ ergs and providing 0.7 $M_{\odot}$
in iron,
the total feedback energy from Type Ia supernovae
cannot exceed about
$E_{SnIa} \approx (0.3,1.0)\times 10^{62}$ ergs, which are
also less than $\Delta PE(r_v)$ in Table 2.

While AGN feedback dominates cluster energetics,
supernovae may nevertheless
contribute 10-20\% of the total feedback energy.
Nagai et al. (2007a,b) computed a variety of gaseous atmospheres
in clusters including supernovae of all types
but without AGN feedback, as commonly assumed.
In their clusters the baryon fraction and entropy profiles are in
reasonably good agreement
with those observed by Vikhlinin et al. (2006) outside
the central region, $r \gta 0.2r/r_v$.
But central overcooling in $r \lta 0.2r/r_v$
is a serious problem.
At zero redshift in the models of Nagai et al.
about 40\% of the baryons within $r \approx 0.5r_v$
are in the form of a centrally concentrated mass of stars and
cold gas.
The mass of this central concentration of
radiatively cooled baryons causes the gas density and
temperature to peak up near the cluster
centers unlike the observations.
A large amount of AGN feedback energy, similar to
that estimated here, is essential to
remove this overcooling gas before it forms into stars and relocate it
to distant regions of the cluster.
The outward flow of cluster gas that results from
the creation of X-ray cavities is described by
Mathews \& Brighenti (2008) and Mathews (2009).
AGN feedback has just begun
to be included in recent cosmological cluster calculations
where the overcooling problem is greatly alleviated
(Teyssier et al. 2010; Puchwein et al. 2010; McCarthy et al. 2010).

\subsection{Feedback stops cooling flows within the 
cooling radius}

The approximate rate that mass cools in the two composite 
clusters in the absence of feedback 
can be estimated from the bolometric X-ray luminosity 
at the cooling radius,
\begin{equation}
L_x(r_{cool}) = {\dot M}_{cf} {3\over 2}{kT \over \mu m_p} 
= 1.50\times 10^{41} 
\left( {{\dot M}_{cf}(r_{cool}) 
\over M_{\odot}~{\rm yr}^{-1}} \right) T_{keV}
\end{equation}
(in erg s$^{-1}$) where the cooling radius $r_{cool}$ is defined as
that radius at which the local gas cooling time is equal to 
a typical cluster age $t_{cl}\sim 7$ Gyrs.
For cluster (1,2) with 
$r_{cool} = (98, 120)$ kpc  
and
$L_x(r_{cool}) = (1.36,4.96)\times 10^{44}$ 
erg s$^{-1}$ 
(determined with $\rho_{obs}(r)$) we find 
${\dot M}_{cf} \approx (225,465)$ $M_{\odot}$ yr$^{-1}$. 
This is the approximate rate that the cluster gas 
would cool in the absence of feedback energy. 
The ratio for the two clusters ${\dot M}_{cf,1}/{\dot M}_{cf,2} = 0.48$ 
is comparable to but larger than the ratio of total feedback energies,
$\Delta PE_1/\Delta PE_2 \approx 0.29$ for gas initially at $r_v$ 
in the ``no core'' solutions.

As shown in Mathews (2009), 
feedback energy can create cluster outflows 
that balance and arrest cooling inflows caused by radiation losses. 
For this to happen within a strongly radiating cluster core, 
the time-averaged 
rate that cluster gas flows out due to feedback expansion 
must be equal to the average mass inflow rate 
${\dot M}_{cf}$ due to radiative losses.
First imagine the expanding flow of cluster gas due to feedback, 
ignoring energy losses due to radiation. 
In this limiting case 
the mass of gas within any radius $r$ decreases 
because of feedback expansion from $M_{ad}(r)$ to $M_{ob}(r)$ 
as the initial adiabatic
atmosphere evolves to the observed one
during the cluster lifetime $t_{cl}$.
The mean positive feedback mass flow past radius $r$,  
$\Delta M(r)/t_{cl} \equiv [M_{ad}(r) - M_{obs}(r)]/t_{cl}$, 
is plotted in Figure 5 for both composite clusters.
At the other limit when radiative losses are present but 
feedback is absent, 
the rate that gas cools within radius $r$, ${\dot M}_{cf}(r)$ 
is related by equation (8) to the total rate that energy 
is radiated within this radius, $L_x(r)$.
In particular, 
the mass of gas within the cooling radius in the observed 
cluster atmosphere, by definition of $r_{cool}$, 
is expected to flow inward at an average rate 
${\dot M}_{cf}(r_{cool})$. 
Consequently,  
in the presence of both radiative losses and feedback gains,
feedback from the
central black hole is expected to increase until
the time-averaged rate that gas mass is driven outward
past $r_{cool}$ balances the average rate that mass would 
cool and flow inward past $r_{cool}$ over the cluster lifetime.

This balance is shown in Figure 5 
where the total time-averaged 
rate that mass flows out past the cooling radius,
$\Delta M(r_{cool})/t_{cl}$ 
very nearly matches the approximate mean inflow rate 
${\dot M}_{cf}(r_{cool})$
expected (from eqn. 8) at radius $r_{cool}$ during time $t_{cl}$ 
if radiative cooling within this radius is uninhibited by feedback.
Since inflow ${\dot M}_{cf}(r)$ 
(estimated from $L_x(r)$ without feedback) 
and outflow $\Delta M(r)/t_{cl}$ (estimated without radiation losses) 
have opposite signs, the near equality of their magnitudes 
at $r = r_{cool}$ in Figure 5 
indicates that feedback and cooling are balanced within $r_{cool}$. 
A quasi-steady state is established 
with little or no net gas flow across $r_{cool}$, 
although feedback outflow continues in 
$r > r_{cool}$ unaffected by radiation losses.
 
The important agreement between opposing mass flows 
at $r_{cool}$ shown in Figure 5
is a further confirmation of the self-consistency 
our simple feedback estimates. 
Less than 1 percent of the total 
feedback energy is stored as potential energy 
within the cooling radius.
Nevertheless, central black holes -- 
in their over-zealous, over-reaching efforts 
to feed back to the cluster atmosphere the energy acquired from gas  
cooling in their immediate vicinity -- 
drive huge flows of gas out beyond the 
virial radius, but in the process necessarily provide enough 
mass outflow within the relatively small 
cooling radius to shut down the large cooling inflow 
that would otherwise occur in this critical central region.
Although we do not consider the detailed time evolution 
of the initial adiabatic cluster gas profile 
as it transforms into the 
gas density profiles observed today, we imagine that this occurs 
in a quasi-steady manner, as explained above, in which
feedback energy 
is widely distributed as $PE$ in the cluster gas.
By this means 
the feedback outflow always nearly balances the radiative inflow, 
avoiding any large central gas concentration 
(and eventual overdensity due to star formation) or other 
excursions very far from the gas density profiles currently observed.

\subsection{Feedback production of cosmic rays}

It is likely that feedback consists of jets and jet-produced
cavities that are filled mostly with cosmic rays.
If so, it is interesting
to compare the total feedback energy in our composite
clusters, $\sim10^{63}$ ergs, with that expected from other
cluster cosmic ray sources.
For simplicity, consider proton cosmic rays that
have cluster lifetimes comparable to the cluster age 
and compare the total (proton) cosmic ray energy from feedback 
shocks to the much stronger accretion shock 
that produced the underlying entropy gradients in 
cluster atmospheres. 
From Table 2 the total potential energy of
gas within the virial radius in clusters 1 and 2 is
$|PE| \sim 10^{64}$ ergs so the total thermal energy is
$E_{th} \sim 0.5\times 10^{64}$ ergs by the virial theorem.
The virial temperature and entropy in the cluster gas
are acquired in the accretion shock as the cluster formed.
Typically, about 10\% of the shock energy is converted to cosmic rays.
so the total energy in cosmic rays created as the cluster formed is
$E_{cr} \sim 0.1E_{th} \sim 0.5\times 10^{63}$ ergs.
Therefore, the cosmic ray energy from feedback and cluster accretion 
are comparable.

The thermal energy profiles in our adiabatic,
pre-feedback cluster models (upper panels in Figs. 1-4)
change very little after receiving the
enormous feedback energies we consider.
In our scenario almost all of the thermal energy created
when feedback energy is initially deposited subsequently
becomes potential energy as the cluster gas expands;
this is the energy evolution described
in Mathews \& Brighenti (2008).
More likely, most of the feedback energy is supplied
from the central black holes as jets of supra-thermal cosmic
rays that drive the shocks that increase the cluster gas entropy.
When feedback is mostly in the form of cosmic rays,
the net thermal energy of the cluster gas is subsequently
reduced by cluster expansion more than it is
initially increased by
shock waves and the net effect of
cavity production is to cool, not heat, the cluster gas
(Mathews \& Brighenti 2008).
AGN feedback energy and cavity production 
in cosmological calculations are assumed to be
in the form of ultra-hot thermal gas.
This injection of thermal
energy results in a small net increase in the total
thermal energy of the cluster gas even after it has
fully expanded.
Since we do not explicitly consider the non-thermal cosmic ray
component here, a very small increase (rather than a decrease)
in the post-feedback temperature profiles appears in Figures 1-4.
In either case, any small change in the thermal energy
of the cluster gas can be neglected
in our estimates of the total feedback energy.

\subsection{Locations of feedback energy deposition and storage}

It is important to recognize the distinction between 
the location in cluster gas where entropy-increasing 
feedback energy is deposited and where it is ultimately 
stored as potential energy which is what we address here. 
Evidently most feedback heating occurs 
in shock waves moving away from advancing jets and expanding 
X-ray cavities. 
X-ray observations of these shocks generally indicate modest 
Mach numbers, ${\cal M} \la 2$, in which the gas entropy 
is increased only by $\la 1.2$. 
The lower density of recently shock-heated gas results in  
an adiabatic, buoyant 
outward flow in the cluster that stops when the entropy 
of the heated region matches that of the ambient atmosphere.
However, the entropy increases with cluster radius 
as $S \propto r^{1.2}$ mostly due to dissipation acquired 
during cosmic accretion. 
Consequently, gas heated with a ${\cal M} \sim 2$ 
shock can only rise by a factor of $\sim 1.2$ in cluster radius 
where its feedback energy is stored as potential energy.
Conversely, if the reduced gas fraction 
observed at $0.5r_v \sim 1$ Mpc 
were due to shock heating events at 50 kpc (where X-ray 
cavities can be observed), the entropy difference 
$\Delta S \sim 36$ would require shocks with Mach numbers 
${\cal M} \sim 17$ in which the post-shock temperature 
would be increased by about 95. 
Shocks this strong have not been observed in cluster gas.
Furthermore, it is likely that there is not enough cluster gas 
within 50 kpc which, when continuously heated this much with a 
filling factor of unity, can buoyantly  
supply the much larger mass of high entropy, 
feedback heated gas observed at 1 Mpc. 
In addition,
mixing instabilities are likely to dilute the entropy of 
heated gas during their long buoyant journey to $r \sim 1$ Mpc.
Clearly, long range buoyant outflow is a losing proposition.
The conclusion we draw from this is that feedback energy 
stored at some radius always exceeds the radius where the 
energy was initially deposited, but the difference
between these two radii is unlikely to be very large.

\section{Black Hole Accretion or Spin Energy?}

The time-averaged mechanical feedback power $L_{fb}$ required to lift 
the adiabatic cluster atmospheres beyond the virial radius $r_v$ can 
be estimated from  
$L_{fb} \sim \Delta PE / t_{cl}$ where
$t_{cl} = 7$ Gyrs is a typical age for large clusters.
For composite cluster (1,2) the feedback luminosity 
with and without initial cores is 
$L_{fb} \sim (0.38,1.32)\times 10^{46}$ and
$L_{fb} \sim (0.44,1.50)\times 10^{46}$ erg s$^{-1}$
respectively. 
These luminosities are comparable to those of 
powerful quasars continuously active during
the cluster lifetime.

What are the implications of these enormous feedback powers 
for the mass of cluster-centered black holes? 
The conversion of accretion energy into black hole 
mass is more straightforward for luminous quasars 
in which the (thin) accretion disk can be directly observed.
In this case, 
the total bolometric feedback energy radiated 
by an active black hole is related to the 
mass of the 
black hole by $E_{rad} = \epsilon_r M_{bh}c^2$ with
$\epsilon_r \sim 0.1$
(e.g. Davis \& Laor 2010). 
While low redshift cluster-centered black holes generally have 
bolometric luminosities very much less than 
the estimated $L_{fb}$, 
it is unclear if this also applies at higher redshift 
when most of the feedback energy may have been created.

Suppose we adopt a mechanical feedback efficiency
$\epsilon_{mfb}$ such that the change in cluster gas potential
energy is related to mass accretion by
\begin{equation}
\Delta PE = \epsilon_{mfb} \Delta M_{bh}c^2.
\end{equation}
If we identify the total mass accreted with the mass
of the black hole, $\Delta M_{bh} \approx M_{bh}$,
and assume $\epsilon_{mfb} \approx 0.1$,
the resulting black hole masses for cluster 1,
$M_{bh} \approx 4.6 - 5.4\times 10^9$ $M_{\odot}$,
are similar to
those observed in cluster-centered galaxies,
e.g. $M_{bh} \approx 6.4\times 10^9$ $M_{\odot}$
in M87 at the center of the Virgo cluster
(Gebhardt \& Thomas 2009).
However, the corresponding black hole masses for the more massive 
cluster 2, 
$M_{bh} \approx 1.5 - 1.6\times 10^{10}$ $M_{\odot}$,
exceed those observed. 
This large black hole mass cannot be reduced by 
invoking the energy contributed to the cluster gas from 
accreting black holes in bulges of non-central satellite galaxies 
in the cluster. 
Almost all observed X-ray cavities and shock waves are associated 
with cluster-centered black holes,
not those in orbiting satellite galaxies. 
Moreover, the cluster center is where the virialized cluster gas 
is densest and has the shortest radiative cooling time.
However, for cluster masses $M_v \gta 10^{14}$ $M_{\odot}$, 
such as we consider here, the mass of the central 
galaxy (and therefore also its black 
hole) increases very little with increasing cluster virial mass $M_v$ 
(Lin \& Mohr 2004).
The mass of the central black hole in 
cluster A478 (in composite cluster 2)
estimated from the bulge 
luminosity of the cluster-centered galaxy, 
$5.8\times 10^9$ $M_{\odot}$ 
(McNamara, Rohanizadegan, \& Nulsen 2011), 
is considerably less than the mass estimated 
from $M_{bh} \approx \Delta PE(r_v)/(\epsilon_{mfb}c^2) 
\approx 1.6\times 10^{10}$ $M_{\odot}$ 
with $\epsilon_{mfb} = 0.1$ for the cored cluster 2. 
Finally, the mechanical accretion efficiency 
$\epsilon_{mfb}$ cannot be much larger than 
0.1 and is likely to be much smaller.

This difficulty can be alleviated if a substantial 
fraction of the required mechanical feedback energy 
is supplied by the rotational energy of central 
black holes which, when magnetically coupled to accretion
disks, may form powerful feedback jets 
(Blandford \& Znajaek 1977).
The maximum available energy from a rotating black hole, 
$0.29M_{bh}c^2 = 5.3\times 10^{62}[M_{bh}/(10^9 M_{\odot})]$ 
erg, is large enough to account for the gas 
depletion in both clusters 1 and 2. 
Recent models of the cosmological co-evolution of galaxies 
and their black holes are consistent with the assumption 
that ``radio mode''
jet feedback derives from black hole spin 
and low-luminosity, advection-dominated central accretion
(Sikora, Stawarz, \& Lasota, 2007; 
Fanidakis et al. 2011).
There is considerable evidence for mechanical
outflow from active
galaxies and quasars (e.g. Crenshaw et al. 2003),
but most relevant to our feedback estimates here are powerful
FRII radio sources in luminous quasars
that can transport enormous energies to great
distances in the cluster gas (e.g. Mullin et al. 2008).
Recently Singal et al. (2011) argue that
the evolution of quasar luminosities
at optical and radio frequencies are strongly
correlated since redshift $z \approx 3$
with a significantly higher radio than optical evolution.
In all likelihood, most of the radio-mode feedback energy from
cluster-centered black holes probably occurred at earlier times.
While spin energy is an attractive hypothesis, it must be 
reconciled with the approximately isotropic ejection of 
feedback jet energy currently observed 
in local galaxy clusters such as Virgo and 
Perseus and the rather large number of FRII radio sources 
with multiple hotspots, both of which, on the spin 
hypothesis, would indicate abrupt changes in the massive  
black hole spin axis on timescales that are implausibly 
small.

Throughout this discussion we assume that the huge energy 
required to drive down cluster gas fractions originates 
in (cluster-centered) black holes. 
While other as yet unidentified sources of energy cannot 
be excluded, a cluster-centered energy source seems likely.
The largest bipolar X-ray cavities associated with 
feedback from cluster centers require energies 
$\sim 10^{62}$ ergs 
(McNamara et al. 2005; Guo \& Mathews 2010b) that are 
a significant fraction of the total feedback energies we estimate.
Strong {\it ad hoc} preheating of baryons to a fixed 
adiabat before they 
flow across the cluster virial radius produces a characteristic 
entropy ``floor'' that may be inconsistent with the 
mass of gas observed in smaller group and galaxy potentials. 
To produce the quasi-powerlaw entropy profiles 
observed in clusters (dotted lines in Figs. 1-4) 
the preheated gas would need to be differentially heated 
in a fine-tuned manner, varying with redshift.

\section{Relation Between Gas Fraction and Entropy}

In their recent paper Pratt et al. (2010) plot entropy 
and gas fraction profiles for over 30 nearby clusters 
observed with {\it XMM-Newton}, 
confirming the strong anti-correlation between these two parameters.
When the mean entropy $\langle S(r)\rangle$ increases, 
the gas fraction $\langle f_g\rangle$ decreases and the amplitude 
of this behavior increases with decreasing cluster gas temperature
and $M_v$ as expected. 
But, in an interesting departure from previous work, 
in their Figure 9 Pratt et al. plot a combination of the 
normalized entropy and the integrated, mass-weighted gas fraction  
$[S(r)/S_{500}](f_g(<r)/f_b)^{2/3}$ with $r/r_{500}$, 
showing that the previously discordant profiles for 
$S(r)/S_{500}$ among their 30 clusters 
all collapse almost magically into a single tight 
bundle of correlations right along the 
relation expected for adiabatic baryonic collapse,
$S_{ad}(r)/S_{500} = 1.42(r/r_{500})^{1.1}$ 
for $r \la r_{500}$. 

It is easy to demonstrate a similar result with our 
composite clusters 1 and 2.
The upper panels in Figures 1 and 2 for ``no core'' clusters 
demonstrate that the 
gas temperature varies with cluster radius very much less than 
the observed or adiabatic 
entropy $S(r) = T(r) / \rho(r)^{2/3}$ in the central panels.
The steady increase in entropy with cluster radius 
is due almost entirely to its reciprocal relation to 
decreasing gas density. 
Moreover, the gas densities for the adiabatic and observed clusters 
vary with the total cluster density as  
$\rho_{ad}=f_b\rho_t$ and $\rho_{obs}=f_g\rho_t$. 
Consequently, the adiabatic gas density at any radius 
is related to the observed density by 
$\rho_{ad} = \rho_{obs}(f_b/f_g)$ where $f_b = 0.17$. 
Using this relation, 
the local entropy in our adiabatic, feedback-free clusters can 
be approximately written in terms of the entropy in the observed
cluster at the same radius, 
${\hat S}_{ad}(r) = [T(r) / \rho(r)^{2/3}]_{obs}(f_g/f_b)^{2/3}$, 
where $T(r) \approx T_{ad}(r) \approx T_{obs}(r)$.

Dash-dotted lines in the central panels of Figures 1 and 2
show profiles of ${\hat S}_{ad} = S_{obs}(f_g/f_b)^{2/3}$
which are almost identical to the 
solid line profiles for the adiabatic atmosphere, $S_{ad}(r)$.
This is the result found by Pratt et al.
Although we use the more appropriate local gas mass fraction 
$f_g(r)$ to show that ${\hat S}_{ad} \approx S_{ad}$, 
the gas mass-weighted values of the integrated gas fractions 
$f_g(<r)$ used by Pratt et al. are not greatly different from 
local values $f_g(r)$ because the gas mass increases rather 
rapidly with cluster radius. 
Like Pratt et al. we find that the 
observed entropy profiles for both clusters 
1 and 2 collapse back to the adiabatic profiles when the observed 
density is increased by $f_b/f_g$.

What is the physical significance of this result? 
It is not surprising that the gas densities between observed 
and adiabatic atmospheres are related by 
$\rho_{ad}(r)/\rho_{obs}(r) = f_b/f_g(r)$ 
since this is required by the definition of these two atmospheres.
However, 
the close agreement between ${\hat S}_{ad}(r)$ and $S_{ad}(r)$ is possible
because the gas temperature profiles 
in the observed and adiabatic atmospheres 
(bound by the same virial mass)
are virtually identical in clusters 1 and 2. 
To an excellent approximation,
the observed and adiabatic entropy 
differ locally only because of the differing gas densities.
It is also interesting that
the entropy profiles of all clusters observed by Pratt et al.
-- and possibly all known clusters -- can be modified in this way 
to recover the universal adiabatic profile.
Clusters for which ${\hat S}_{ad}(r) \approx S_{ad}(r)$
are relaxed in the sense
that non-gravitational ``heating'' events that increase
the cluster gas entropy and drive it out in the cluster potential,
lowering the gas fraction,
are either too weak or too old to retain evidence of 
increased gas temperature due to the most recent feedback event.
Conversely, clusters for which ${\hat S}_{ad}(r)$ 
slightly exceeds $S_{ad}(r)$ show evidence of recent 
feedback-related increases in gas temperature and 
${\hat S}_{ad}(r) - S_{ad}(r)$ 
is a measure of the location, energy and age of these feedback events.

\section{CONCLUSIONS}

By considering the difference between gas density profiles in clusters 
created by non-radiating gravitational collapse 
and similar observed clusters,
we estimate the total feedback energy 
received by the gas during the cluster lifetime. 
This estimate is insensitive to the precise time of the feedback 
events, but on average feedback energy 
must be widely distributed in cluster radius.
Individual feedback events produce jets and cavities 
and associated shock waves that heat the cluster gas.
The heated, high entropy gas expands, ultimately leading to 
a expansion of the entire cluster atmosphere 
in which the feedback energy is stored as potential energy.
The final cluster gas entropy profile is increased 
by feedback energy and the gas fraction $f_g(r)$ profile 
is reduced below the cosmic baryonic value $f_b = 0.17$.

For clusters having masses $M_v \gta 10^{14}$ $M_{\odot}$
the total estimated feedback energy required to account 
for the observed depletion of cluster gas and the enhanced 
entropy profile is $\ga10^{63}$ ergs,
considerably in excess of cluster gas energy gained from supernovae 
or lost by radiation. 
Although enormous, this energy exceeds the largest known 
energy released in single, most powerful feedback events, 
$\la 10^{62}$ ergs.
The most likely source of this energy is feedback 
from central black holes in cluster-centered elliptical galaxies.
When averaged over a typical cluster lifetime, about $7$ Gyrs, 
the mean mechanical luminosity, $\sim10^{46}$ ergs s$^{-1}$, 
is comparable to that of powerful quasars. 
Such a large sustained luminosity may require 
energy creation not just from black hole accretion but also 
its spin energy.

Immediately following a feedback energy event, 
we expect a significant local increase in gas temperature 
and thermal energy.
But after a cluster sound-crossing time 
the cluster expands and the temperature returns 
to a rather flat profile near the virial temperature 
required to support the cluster gas.
The sound-crossing time to $r_{500}$ in 
our two composite clusters (1,2)
is $t_{500} = (4.6,3.5)\times 10^8$ yrs.
In view of the insensitivity of the gas temperature 
profile to the feedback energy after time $\sim10^8$ yrs, 
changes in the cluster gas entropy $S(r)=T(r)/\rho(r)^{2/3}$ 
arise almost entirely from changes in the gas density profile. 
After $\sim10^8$ yrs,
transient feedback events that increase the local gas temperature
evolve by expansion into density reductions
as thermal energy converts to potential energy,
retaining approximately the same global $T(r)$.
This explains the strong anti-correlation between excess entropy 
and reduced gas fraction in galaxy groups and clusters 
(Sun et al. 2009).
Inspired by Pratt et al. (2010), we also demonstrate 
that the entropy profile observed in 
any relaxed cluster $S(r)$, when multiplied by 
a factor containing the gas fraction $f_g(r)$,
recovers the universal adiabatic gas entropy profile expected 
in the absence of feedback, 
${\hat S}_{ad}(r) = S(r)[f_b/f_g(r)]^{2/3} \propto r^{1.2}$,
with small deviations related to recent feedback events. 

Most of the feedback energy and entropy are 
delivered to very distant regions in cluster 
hot gas atmospheres, far beyond the cooling radius,
where they have little or no effect on 
reducing the rate that gas cools near the central black hole, 
the presumed source of feedback energy.
In the absence of feedback energy, the observed cluster 
gas is expected to cool at a rate ${\dot M}_{cf}(r_{cool})$ within the 
cooling radius $r_{cool}$, the cluster radius 
at which the cooling time equals the typical cluster age $t_{cl}$.
For our clusters $r_{cool} \approx 100$ kpc.
Cooling flows near the central black hole can be 
greatly reduced or stopped if the mass of cluster gas  
that flows out across $r_{cool}$ 
during the cluster lifetime $t_{cl}$, driven by feedback, 
is approximately equal to the inflowing mass due to 
radiation losses 
${\dot M}_{cf}(r_{cool})t_{cl}$.
For the clusters we consider 
it is gratifying that the cluster gas mass flow in both 
directions at $r_{cool}$ -- out due to feedback 
and in due to radiation losses -- 
are very nearly equal, 
indicating that approximately the right amount of feedback energy 
is delivered to gas at $r<r_{cool}$ during $t_{cl}$ 
to drastically reduce cooling near the black hole 
where little or no cold or cooling gas is observed. 
Nevertheless, only a small fraction of the total feedback energy,
less than 1 percent, is delivered and stored within $r_{cool}$.

Finally, it is significant that the average mechanical 
feedback power $L_{fb} \sim 10^{46}$ erg s$^{-1}$
implied by observed cluster gas fractions 
$f_g < f_b = 0.17$,  
are very substantially higher than estimates of $L_{fb}$ from 
observations of X-ray cavities, $\sim 10^{59}$ ergs  
(Rafferty et al. 2006; McNamara, Rohanizadegan, \& Nulsen, 
2011). 
This discrepancy may be attributed to the difficulty of 
detecting X-ray cavities at distances exceeding about 
50-70 kpc from cluster centers, particularly at higher redshifts, 
where most of the feedback energy is delivered and stored.

\vskip.1in
\acknowledgements
Helpful comments from Ming Sun, Fabrizio Brighenti 
and David Buote are acknowledged. 
Studies of the evolution of hot gas in elliptical galaxies
at UC Santa Cruz are supported by NSF and NASA grants 
for which we are very grateful.


\clearpage

\begin{figure}
\vskip3.in
\centering
\includegraphics[bb=250 216 422 769,scale=.8,angle=270]{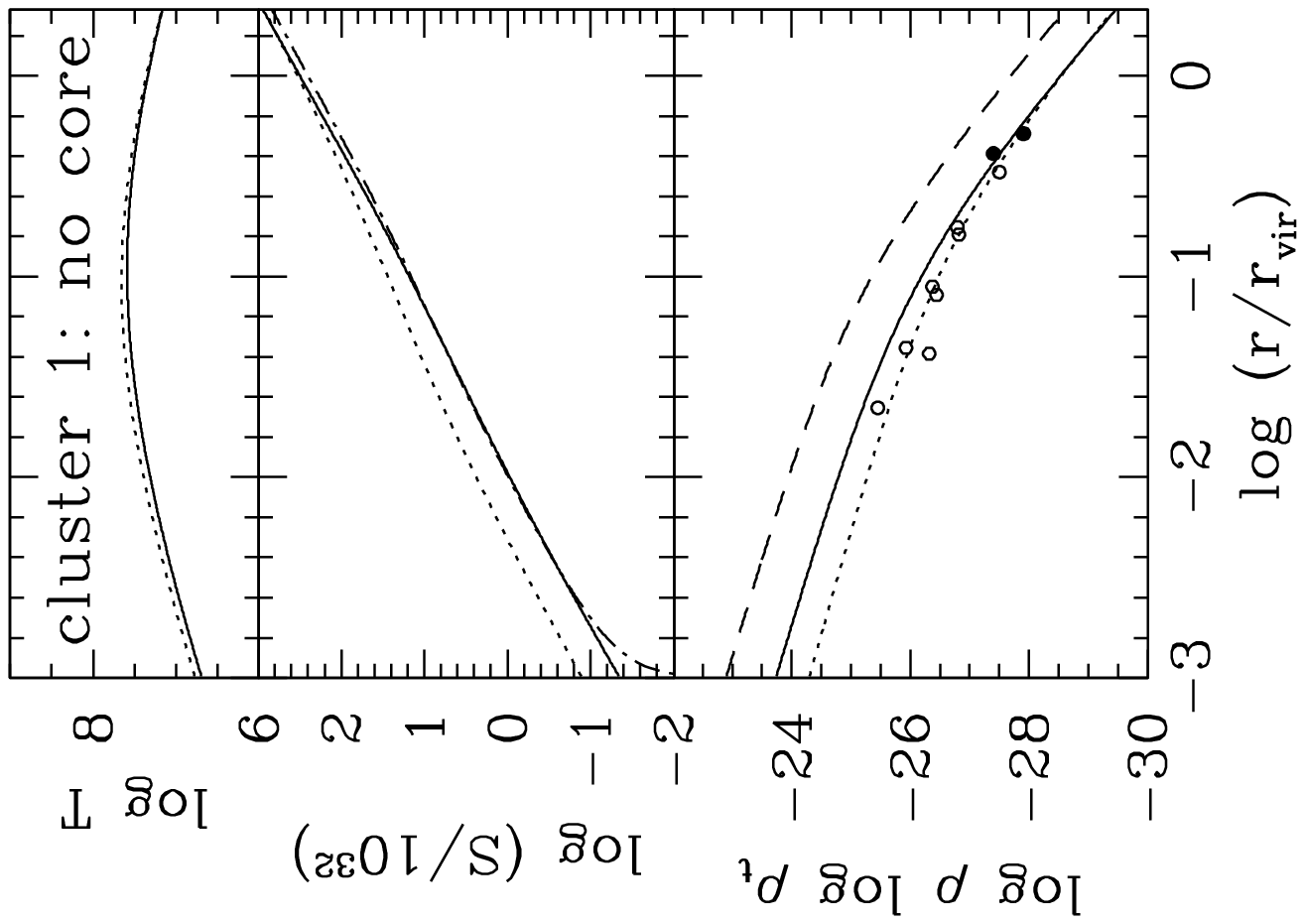}
\vskip1.7in
\caption{
Two equilibrium atmospheres for 
composite cluster 1: a purely adiabatic 
virialized gaseous atmosphere without central core (solid lines)
and an observed atmosphere in a cluster 
of the same mass (dotted lines).
{\it Upper panel}: Gas temperature profiles.
{\it Center panel}: Gas entropy profiles. 
The dash-dotted line shows the adiabatic entropy  
${\hat S}_{ad}$ estimated 
from the observed entropy at the same radius (see Section 4).
{\it Lower panel}: Gas density profiles.  Points show observed gas 
densities for A133 and A383 from Vikhlinin et al (2006) which 
fit with the dotted line. The total 
cluster density profile $\rho_t(r)$ is shown with a dashed line. 
The cooling radius for cluster 1, 98 kpc, corresponds to 
$\log r_{cool}/r_{vir} = -1.30$.
}
\label{f1}
\end{figure}

\clearpage

\begin{figure}
\vskip3.in
\centering
\includegraphics[bb=250 216 422 769,scale=.8,angle=270]{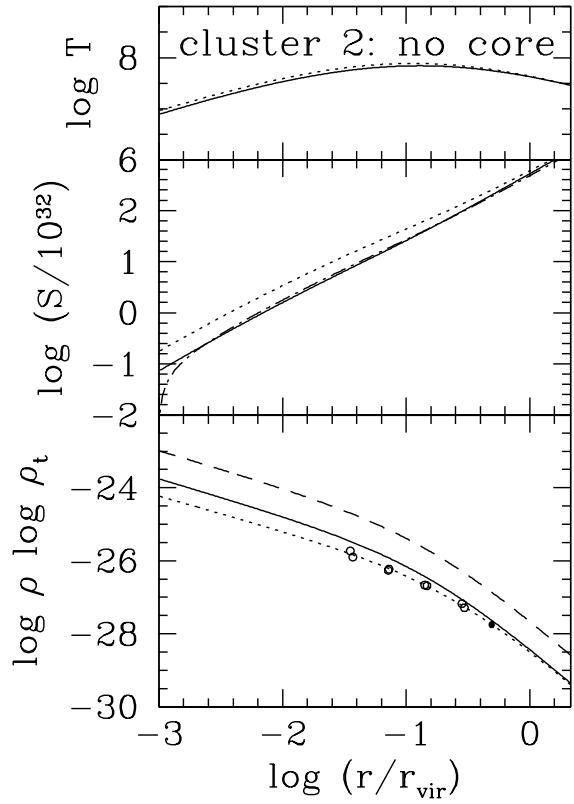}
\vskip1.7in
\caption{
Identical to Fig. 1 but for composite cluster 2 
with observed gas densities for A478 and A1413.
The cooling radius for cluster 2, 120 kpc, corresponds to
$\log r_{cool}/r_{vir} = -1.35$.
}
\label{f2}
\end{figure}

\clearpage

\begin{figure}
\vskip3.in
\centering
\includegraphics[bb=250 216 422 769,scale=.8,angle=270]{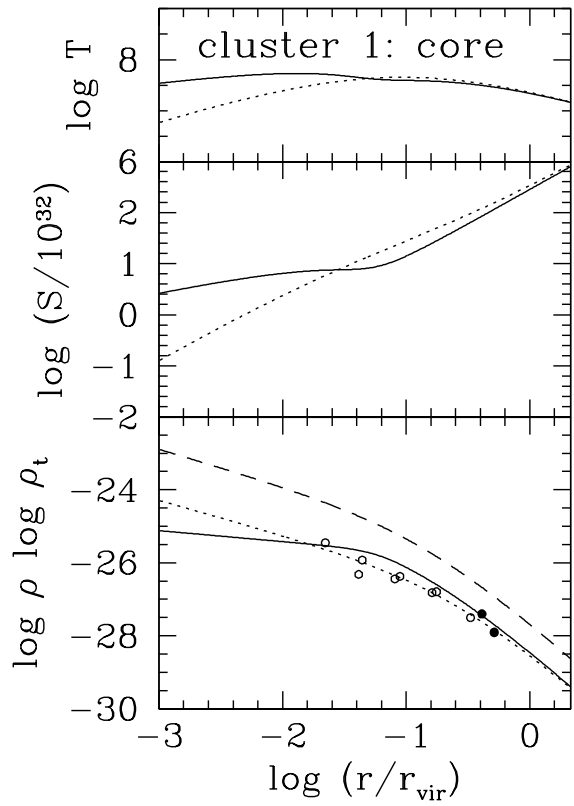}
\vskip1.7in
\caption{
Two equilibrium atmospheres for
composite cluster 1: a purely adiabatic
virialized gaseous atmosphere with central core (solid lines)
and an observed atmosphere of the same mass (dotted lines).
Panels are identical to Fig. 1.
}
\label{f3}
\end{figure}

\clearpage

\begin{figure}
\vskip3.in
\centering
\includegraphics[bb=250 216 422 769,scale=.8,angle=270]{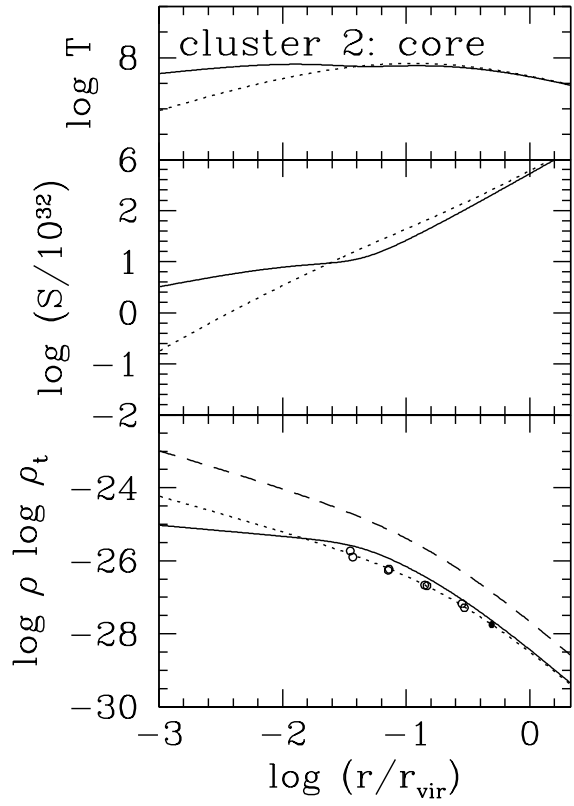}
\vskip1.7in
\caption{
Identical to Fig. 3 but for composite cluster 2 
with observed gas densities for A478 and A1413.
}
\label{f4}
\end{figure}

\clearpage

\begin{figure}
\vskip3.in
\centering
\includegraphics[bb=250 216 422 769,scale=.8,angle=270]{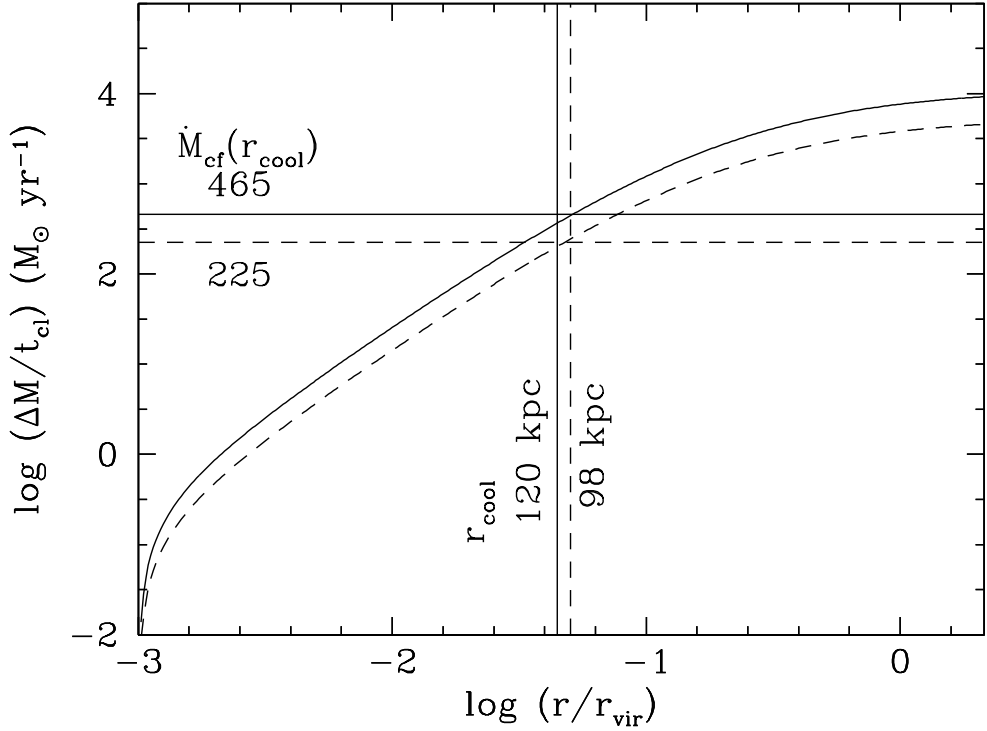}
\vskip1.7in
\caption{
Sloping lines show profiles 
of the approximate time-averaged rate $\Delta M(r)/t_{cl}$
that the local integrated gas mass is lowered due 
to feedback outflows at each (no core) cluster radius 
and $t_{cl} = 7$ Gyrs is the cluster lifetime.
Horizonal lines show the rate ${\dot M}(r_{cool})$ that gas cools 
by radiation as it flows inward across $r_{cool}$ in the absence 
of feedback. 
The cooling radius $r_{cool}$ is marked with vertical lines.
(Dashed,solid) lines refer to composite cluster (1,2).
For both clusters 
the mean rate of mass outflow at the cooling radius 
$\Delta M(r_{cool})/t_{cl}$ 
is very close to the mass inflow rate at this radius 
${\dot M}_{cf}(r_{cool})$ due to radiative losses. 
This near equality is a necessary condition for feedback energy to arrest 
cooling flows.
}
\label{f5}
\end{figure}

\clearpage

\vskip1.in

\begin{deluxetable}{lcrrcr}
\tabletypesize{\scriptsize}
\tablecolumns{6}
\tablewidth{8.5cm}
\tablecaption{TWO COMPOSITE GALAXY CLUSTERS\tablenotemark{a}}
\tablehead{
\colhead{cluster} &
\colhead{$M_{500}$\tablenotemark{b}} &
\colhead{$r_{500}$\tablenotemark{b}} &
\colhead{$T_{mg}$\tablenotemark{c}} &
\colhead{$M_v$\tablenotemark{d}} &
\colhead{$r_v$\tablenotemark{d}} \cr
\colhead{} &
\colhead{($10^{14}$ $M_{\odot}$)} &
\colhead{(kpc)} &
\colhead{(keV)} &
\colhead{($10^{14}$ $M_{\odot}$)} &
\colhead{(kpc)} \cr
}
\startdata
A133 & 3.17 & 1007 & 3.67 & \nodata & \nodata \cr
A383 & 3.06 &  944 & 4.37 & \nodata & \nodata \cr
cluster 1 & 3.11 &  974 & 4.02 & 4.34 & 1950 \cr
     &      &       &      &         &       \cr
A478 & 7.68 &  1337 & 7.36 & \nodata & \nodata \cr
A1413 & 7.57 & 1299 & 6.81 & \nodata & \nodata \cr
cluster 2 & 7.63 & 1319 & 7.08 & 11.3 & 2682 \cr
\enddata
\tablenotetext{a}{Cluster data from Vikhlinin et al. (2006).}
\tablenotetext{b}{Cluster mass and radius at overdensity
$\Delta = 500$.}
\tablenotetext{c}{Density weighted mean temperature of
cluster gas.}
\tablenotetext{d}{Cluster mass and radius at virial radius.}
\end{deluxetable}

\clearpage

\vskip1.in

\begin{deluxetable}{lcccc}
\tabletypesize{\scriptsize}
\tablecolumns{5}
\tablewidth{7.cm}
\tablecaption{COMPOSITE GALAXY CLUSTER ATMOSPHERES}
\tablehead{
\colhead{} &
\colhead{$r/r_v$} &
\colhead{$M_{g}(r)$\tablenotemark{a}} &
\colhead{$|PE|$} &
\colhead{$\Delta |PE|$} \\
\colhead{} &
\colhead{} &
\colhead{($10^{13}$ $M_{\odot}$)} &
\colhead{($10^{62}$ ergs)} &
\colhead{($10^{62}$ ergs)} \\
}
\startdata
\cutinhead{{\sc cluster 1: no core}} 
ad\tablenotemark{b} & 1.000 & 7.382 & 42.06 & \nodata \\
ob\tablenotemark{b} & 1.862 & 7.383 & 32.37 & 9.69 \\
ad & 0.501 & 4.459 & 30.38 & \nodata \\
ob & 0.933 & 4.453 & 24.17 & 6.21 \\
ad & 0.269 & 2.486 & 19.57 & \nodata \\
ob & 0.501 & 2.476 & 16.15 & 3.42 \\
\cutinhead{{\sc cluster 2: no core}} 
ad & 1.000 & 19.22 & 204.6 & \nodata \\
ob & 1.549 & 19.20 & 171.6 & 33.0 \\
ad & 0.501 & 11.23 & 142.4 & \nodata \\
ob & 0.776 & 11.19 & 122.4 & 20.0 \\
ad & 0.324 & 7.362 & 103.2 & \nodata \\
ob & 0.501 & 7.317 & 90.26 & 12.94 \\
\cutinhead{{\sc cluster 1: core}}
ad & 1.000 & 7.221 & 40.46 & \nodata \\
ob & 1.820 & 7.275 & 32.12 & 8.34  \\
ad & 0.501 & 4.299 & 23.79 & \nodata \\
ob & 0.891 & 4.284 & 23.58 & 5.21 \\
ad & 0.282 & 2.448 & 18.74 & \nodata \\
ob & 0.501 & 2.476 & 16.15 & 2.59 \\
\cutinhead{{\sc cluster 2: core}} 
ad & 1.000 & 19.00 & 200.7 & \nodata \\
ob & 1.549 & 19.19 & 171.6 & 29.1 \\
ad & 0.501 & 11.01 & 138.5 & \nodata \\
ob & 0.758 & 10.96 & 120.7 & 17.8 \\
ad & 0.331 & 7.325 & 101.3 & \nodata \\
ob & 0.501 & 7.317 & 90.26 & 11.0 \\
\tablenotetext{a}{For each cluster/core combination, the 
integrated gas mass 
is chosen to be nearly the same for each pair of 
``ob'' and ``ad'' solutions.} 
\tablenotetext{b}{``ad'' and ``ob'' designate cluster atmospheres 
based on the adiabatic and observed gas density profiles.}
\enddata
\end{deluxetable}

\end{document}